\documentstyle[prl,twocolumn,aps,epsf]{revtex}

\begin{document}

\twocolumn[\hsize\textwidth\columnwidth\hsize\csname @twocolumnfalse\endcsname

\title{Optical Conductivity of Ferromagnetic Semiconductors}

\author{E. H. Hwang$^{(1)}$, A. J. Millis$^{(2)}$, S. Das Sarma$^{(1)}$}

\address{$^{(1)}$ Condensed Matter Theory Center,
Department of Physics, University of Maryland, College Park, 
MD 20742\\
$^{(2)}$ Department of Physics and Astronomy, Columbia University,
538 W 120th St., New York, NY 10027}

\maketitle

\begin{abstract}
The dynamical mean field method is used to calculate the frequency and 
temperature dependent  conductivity of dilute magnetic semiconductors. 
Characteristic qualitative
features are found distinguishing weak, intermediate, and 
strong carrier-spin coupling and allowing quantitative determination of 
important parameters defining the underlying ferromagnetic mechanism. 


\end{abstract}
\vspace{.25cm}
]

\narrowtext 

Observations of ferromagnetism in dilute magnetic semiconductors 
(DMS) \cite{Ohno98,Ohno96,Matsukura98,Reed01}
with transition temperatures as high as 110K
(GaMnAs at $5\%$ Mn \cite{Ohno98,Ohno96,Matsukura98}) 
or even above room temperature (GaMnN, GaMnP\cite{Reed01}) 
have renewed interest in these remarkable systems 
\cite{Akai98,Dietl00,Litvinov01,Dietl02,Bhatt00,Nagaev96,Chattopadhyay01a},
in part because of possible `spintronic' applications \cite{dassarma01}. 
Several crucial 
scientific questions have emerged, most notably what parameters control 
the magnitude of the magnetic transition temperature and how these may 
be optimized.
There is also an urgent need for experimental diagnostics, both for 
determining fundamental parameters such as the strength of the 
carrier-spin coupling, and for determining sample-specific parameters such 
as the carrier density and the degree of spin polarization of the mobile 
carriers. In this Letter we present theoretical calculations 
showing that measurements of the optical conductivity $\sigma(\omega,T)$ 
can be of great help in answering these questions. 
Optical conductivity measurements have proven useful in 
understanding the physics of the colossal magnetoresistance (CMR) 
manganites \cite{Quijada98}
where carrier-spin coupling is also crucial \cite{CMRReview}. 

We find that the DMS optical conductivity 
exhibits novel features not found in the CMR.
To perform our calculations 
we use a recently developed non-perturbative method, the `dynamical mean 
field theory' (DMFT) \cite{Kotliar} 
to calculate $\sigma(\omega,T)$ for the generally 
accepted model of dilute magnetic semiconductors. 
A non-perturbative method is needed because the crucial physics involves 
bound-state formation and other aspects of intermediate carrier-spin 
couplings not accessible to perturbative methods.
The system sizes required for direct numerical simulations 
for this problem are 
impractically large, and obtaining 
accurate dynamical information from numeric is not easy.
The DMFT method can handle dynamics easily.
Our results display striking, sometimes, counterintuitive, dependence 
of conductivity on carrier-spin coupling, carrier density, 
and temperature revealing key features of the underlying physics.
Experimental observation of our predictions should lead to crucial 
information about bound state formation and impurity band physics in 
this problem.

It is generally believed \cite{Chattopadhyay01a} that DMS are described 
by the generalized Kondo lattice model
\begin{eqnarray}
H=H_{host} &-& \sum_{i,\alpha,\beta}J\hat{\bf S}_{i}\cdot \psi _{\alpha }^{
\dag}(R_i)\vec{\sigma}_{\alpha \beta }
\psi_{\beta }(R_i) \nonumber \\
&+& W \psi^{\dag}_{\alpha }(R_i)
\psi_{\alpha }(R_i),
\end{eqnarray}
where $H_{host}$ describes carrier propagation in the host 
semiconductor and the second (magnetic) 
term describes coupling of the carriers to 
an array of (impurity, e.g. Mn) spins at positions $R_i$. 
The coupling has two sources: a spin-spin coupling and a potential 
scattering.
Here we absorb the magnitude of the 
impurity spin into the coupling $J$ (which we take to be positive), 
and represent the spin 
direction by the unit vector $\hat{\bf S}$. 
For simplicity we consider a host material with a single 
non-degenerate band; the extension to the multiband case relevant,
for example, to the GaAs valance band 
is straightforward, involves no new features, and will be 
presented elsewhere. We therefore write
\begin{equation}
H_{host}= \sum_{\alpha}
\int d^3x \psi _{\alpha }^{+}(x)\frac{\nabla ^{2}}{2m}
\psi _{\alpha }(x) + V_{R}(x)\psi _{\alpha }^{+}(x)\psi _{\alpha }(x),
\end{equation}
where $V_R$ is a random potential arising from non-magnetic defects 
in the material. 

The crucial physical issues are revealed by the consideration of a 
ferromagnetic state in which all impurity spins $S_i$ are aligned, 
say, in the $z$ direction. Then the carriers with spin parallel to 
$S_i$ feel a potential $-J+W$ on each magnetic impurity site and
anti-parallel carriers feel a potential $J+W$. These potentials 
self-consistently
rearrange the electronic structure. The spin-dependent part of this 
rearrangement provides the energy gain which stabilizes the 
ferromagnetic state. The key physics issue is, evidently, whether the 
potential $W\mp J$ is weak (so its effect on carriers near the lower 
band edge is simply a scattering phase shift) 
or strong (so only majority spin or perhaps both 
species of carriers are confined into spin-polarized impurity bands). 
Recent density functional supercell calculations \cite{Sanvito01}
suggest that in GaMnAs
$-J+W$ is close to the critical value for bound state formation
for the majority spin systems.
Then as temperature is increased, the Mn spins disorder and it is 
natural to ask how this physics changes. We shall see that all of 
this behavior is clearly revealed in the optical conductivity,
which can therefore be used experimentally to probe the qualitative 
nature of the magnetic coupling.

To compute the conductivity we write $\nabla \rightarrow (\nabla-ieA/c)$
and apply the usual Kubo formula. To evaluate the properties of $H$ we 
employ the DMFT approximation, which amounts to assuming
that the self energy is momentum independent: $\Sigma(p,\omega) 
\rightarrow \Sigma(\omega)$, so that the Green function $G(p,\omega)$
corresponding to $H$ is $G(p,\omega) = [\omega-p^2/2m -
\Sigma(\omega)]^{-1}$. The self energy is given in terms of 
the solution of an impurity problem \cite{Kotliar}. 
The impurity problem requires a momentum cutoff, arising physically from 
the electron band-width. We impose the cutoff by assuming a semicircular 
density of states $D(\epsilon) = \sqrt{4t^2-\epsilon^2}/2\pi t$ with 
$t=(2\pi)^{2/3}/mb$ and $b^3$ the volume per formula unit.
The parameter $t$ is chosen
to correctly reproduce the band edge density of states.
Within this approximation the real part of the 
conductivity is given by
\begin{eqnarray}
\sigma(\Omega,T)&=&\int \frac{d^3p}{(2\pi)^3} \left ( 
\frac{p\cos\theta}{m} \right )^2
\int \frac{d\omega }{\pi }\frac{\left[ f(\omega )-f(\omega +\Omega
)\right] }{\Omega }\nonumber \\
&\times& {\rm Im} G(p ,\omega ) {\rm Im} G(p ,\omega +\Omega),
\label{sigxx}
\end{eqnarray}
Our approximation for the density of states implies $\int {d^3p}/(2\pi)^3
(p\cos\theta/m)^2 \rightarrow \int d\varepsilon 
D(\varepsilon)\Phi(\varepsilon)$ with $\Phi(\varepsilon)
=(4t^2-\varepsilon^2)/3$ \cite{Chattopadhyay00}.

The main panels of Fig. 1 show the evolution of the conductivity with
temperature for two couplings; intermediate-weak ($J=1$, Fig. 1(a)) and 
intermediate-strong ($J=1.5$, Fig. 1(b)); the insets show the majority-spin 
densities of states. Consider first the $T=T_c$ curves (solid lines),
where $T_c$ is the ferromagnetic transition temperature
\cite{Chattopadhyay01a}.
The $J=1$ curve has approximately the Drude form expected for carriers 
scattering off random impurities (a closer examination reveals minor 
differences due to density of states variations near the band edge). 
In the $J=1.5$ case the density of states plot shows that an impurity 
band is formed and the corresponding conductivity has two structures: 
a low-frequency quasi-Drude peak corresponding to motion within the 
impurity band and a higher frequency peak corresponding to excitations 
from the impurity band to the main band. We observe that the 
frequency of the upper peak does not directly give 
the separation between the impurity band Fermi level and the 
conduction band, because the vanishing of the velocity and density 
of states at the band edge means that the lowest unoccupied state is
not optically active.

We now turn to the temperature dependence. For the $J=1.5$ curves we 
see that as $T$ is decreased, the dc conductivity changes only very 
slightly, whereas the width of the low frequency peak increases and 
the high frequency peak moves up in energy and decreases in oscillator 
strength. The increase in energy of the higher frequency peak may be 
understood from the density of states curves, which show a weak 
blue-shift of the 
conduction band edge and a broadening of the impurity 
band. The counter-intuitive broadening (as $T$ is decreased) of the 
lower frequency peak arises 
because, as the spins order, the binding of the carriers to the 
impurity spins 

\begin{figure}
\epsfysize=2.3in
\centerline{\epsffile{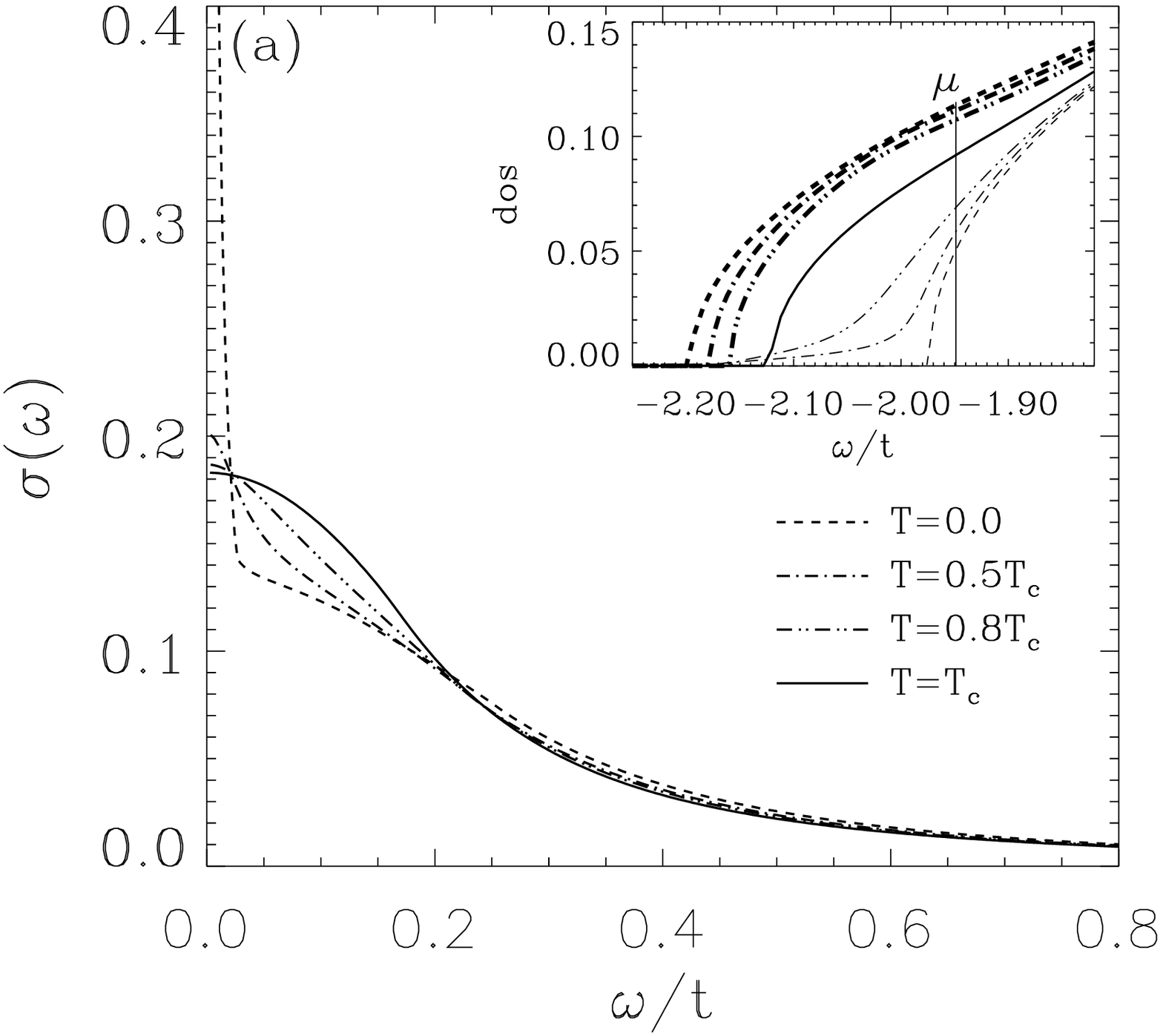}}
\epsfysize=2.3in
\centerline{\epsffile{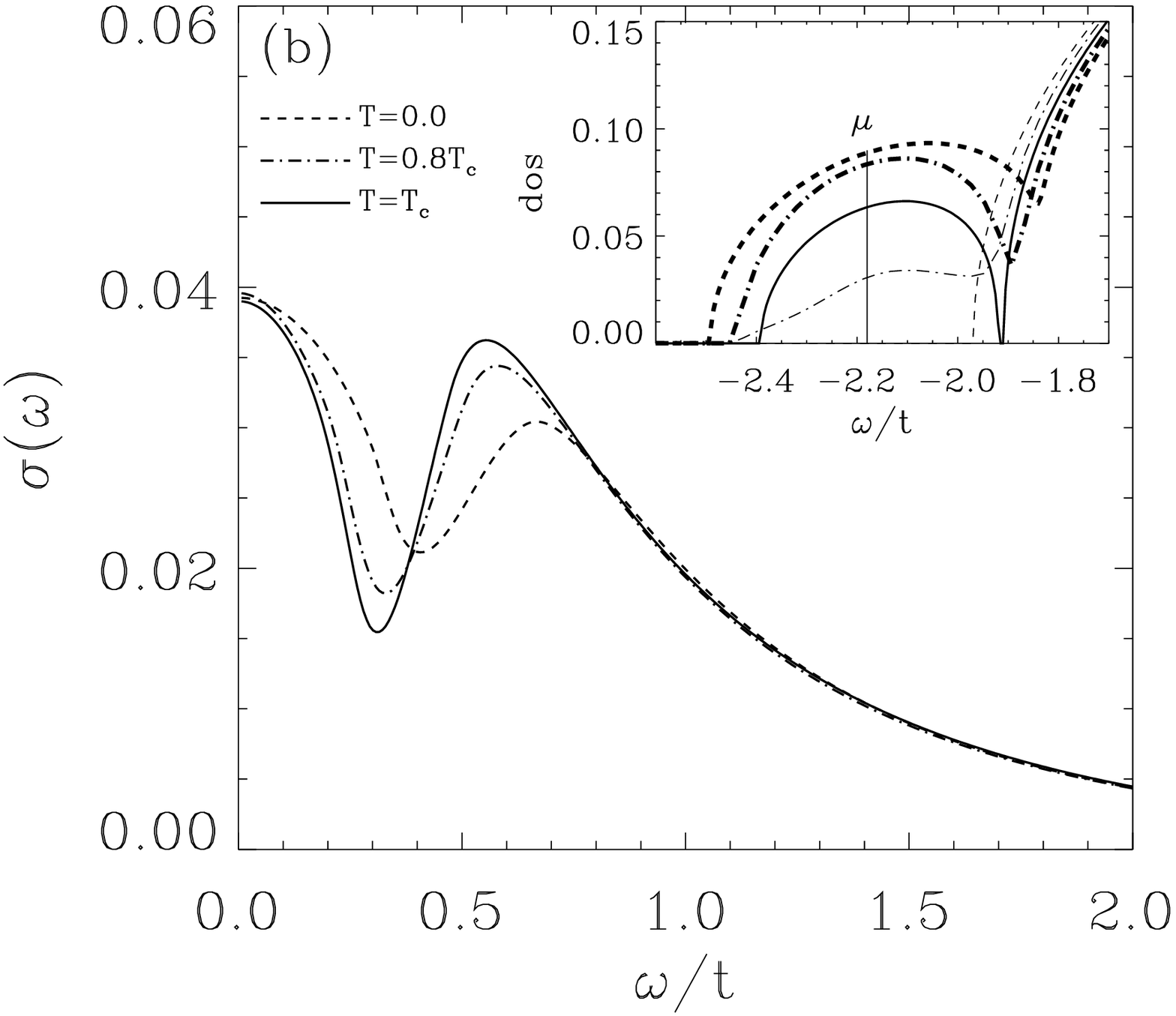}}
\caption{
Temperature dependence of 
optical conductivity for (a) $J=1.0t$, $x=0.05$, $n=0.02$ with
$T_c=5.95\times10^{-3} t/k_B$, and (b)
$J=1.5t$, $x=0.05$, $n=0.02$ with $T_c=1.24\times10^{-2} t/k_B$.
Insets show the density of states for both majority spin 
(thick curves) and minority spins (thin curves). The vertical lines show
the zero temperature chemical potentials.}
\end{figure}

\noindent
increases (as seen from the increase in separation 
between the chemical potential and the edge of the main band) 
corresponding to an increase in the basic scattering rate. 
The weak $T$-dependence of the dc conductivity occurs 
because the increase in scattering rate is compensated 
by another effect.
Because the impurity band is spin-polarized carriers
which are bound to impurity 
site must have spins parallel to impurity spin. Thus, 
as the spins order ferromagnetically, 
basic ability of carriers to move 
in the impurity band increases. This physics is familiar from the 
CMR materials \cite{CMRReview} and corresponds to an increase in 
conduction band oscillator strength.

Consider next the $J=1$ curve, where 
two effects occur 
as $T$ decreases. 
First, the main quasi-Lorentzian peak decreases slightly in 
amplitude and increases slightly in half-width. The increase in width 
is due to increased carrier-spin coupling as mentioned above. 
Second, a new narrow peak appears.
As can be seen from the inset in 
Fig. 1(a), at 
$T=0$ for this carrier concentration, the minority spin-band is slightly 
occupied and the sharply

\begin{figure}
\epsfysize=2.2in
\centerline{\epsffile{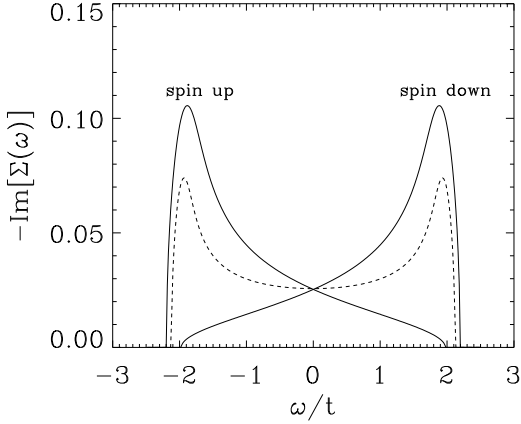}}
\caption{Imaginary part of the self energy as a function of the frequency
for $J=1.0t$ and $x=0.05$.
Solid (dashed) curves show the self energy at $T=0$ ($T=T_c$).}

\end{figure}

\noindent
peaked conductivity occurs because 
this small density of carriers is very weakly scattered. 

To understand the small 
scattering rate it is helpful to consider the Born approximation, 
which would lead to Im$[\Sigma]$ $\sim x N(E_F)J^2$ with $x$ the magnetic 
impurity concentration, $N(E_F)$ the density of states, and $J$ the 
carrier-spin coupling. The combination of the small value of $x$ and 
the small density of states at the band edge leads to a small scattering 
rate (in the Born approximation). Now consider corrections to the Born 
approximation. For the majority spin band, increase of $J$ leads to 
the formation of a spin-polarized bound state, so the corrections to 
the Born approximation must be such as to strongly increase the effective 
scattering rate. On the other hand, for minority spin carriers the 
increase of $J$ leads to an antibound state at the top of the 
band, so that at the physically relevant lower band edge, 
the corrections must be such as to decrease the effective scattering 
rate. Quantitatively, these effects are quite large. Fig. 2, for example, 
shows the calculated imaginary part of the $T=0$ majority and minority 
spin self energy for $J=1$.

We now consider the density dependence of the conductivity. Fig. 3(a) 
shows the evolution of $\sigma(\omega, T=0)$ with carrier density for 
$J=1$. At very 
low density ($n=0.01$) the minority spin band is 
unoccupied and the behavior associated with the majority spin band is 
observed. A sharp peak occurs when the minority band begins to be 
populated. Fig. 3(b) shows the evolution of $\sigma(\omega,T=0)$ with 
$n$ for the `impurity-band' case $J=1.5$. At very 
small $n$ the 
conductivity is dominated by the `impurity band' contribution, with a 
relatively weak feature corresponding to 
transitions from the impurity band to 
the main band. As the carrier density is increased the 
band oscillator strength in transitions between the impurity and main band
increases dramatically, both in absolute
terms and relative to the intra-impurity band transitions.
When the Fermi level crosses into the minority spin band 
(note that the minority spin impurity band is at the upper band edge; 
here we have only main-band states) an

\begin{figure}
\epsfysize=2.3in
\centerline{\epsffile{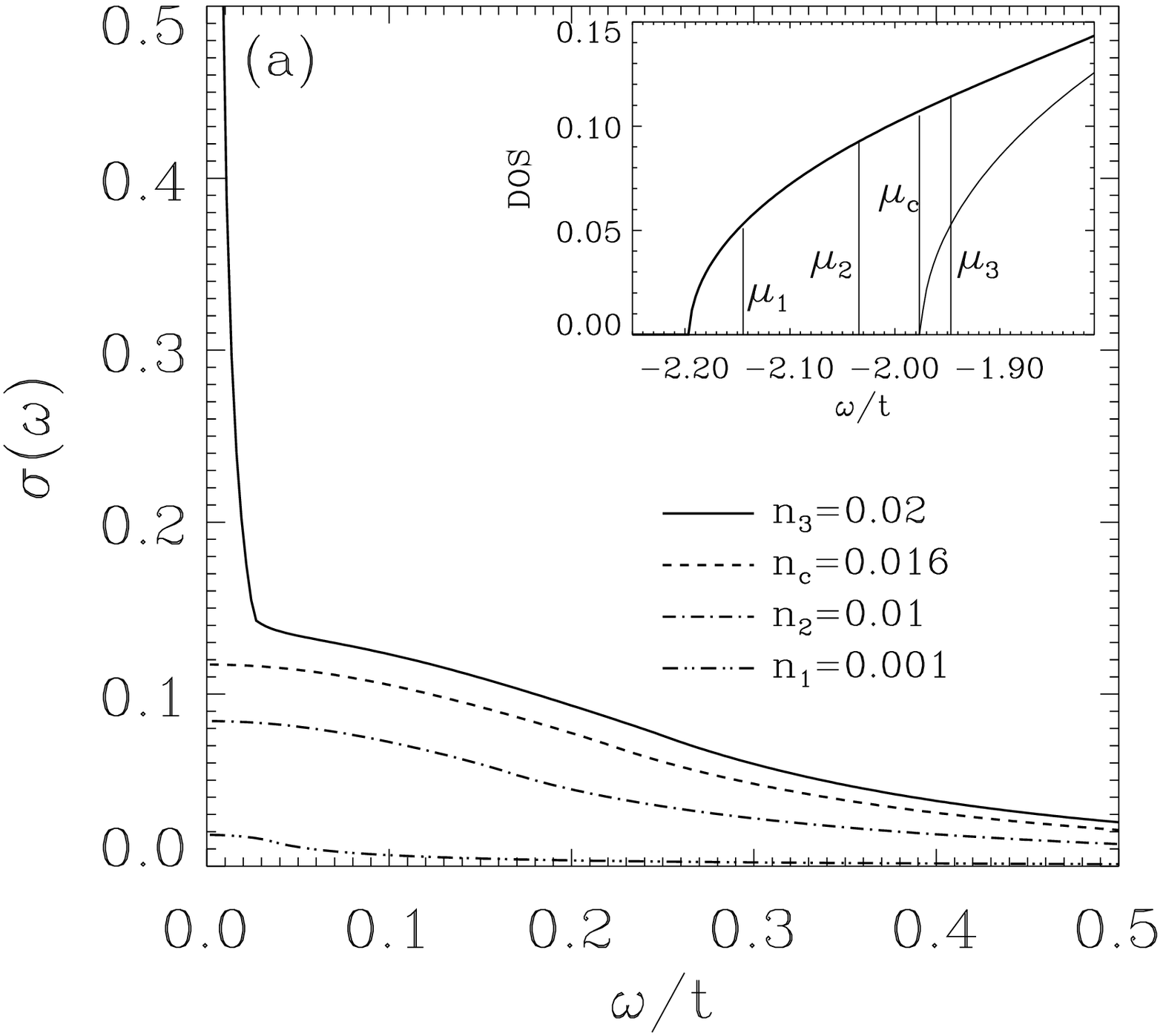}}
\epsfysize=2.3in
\centerline{\epsffile{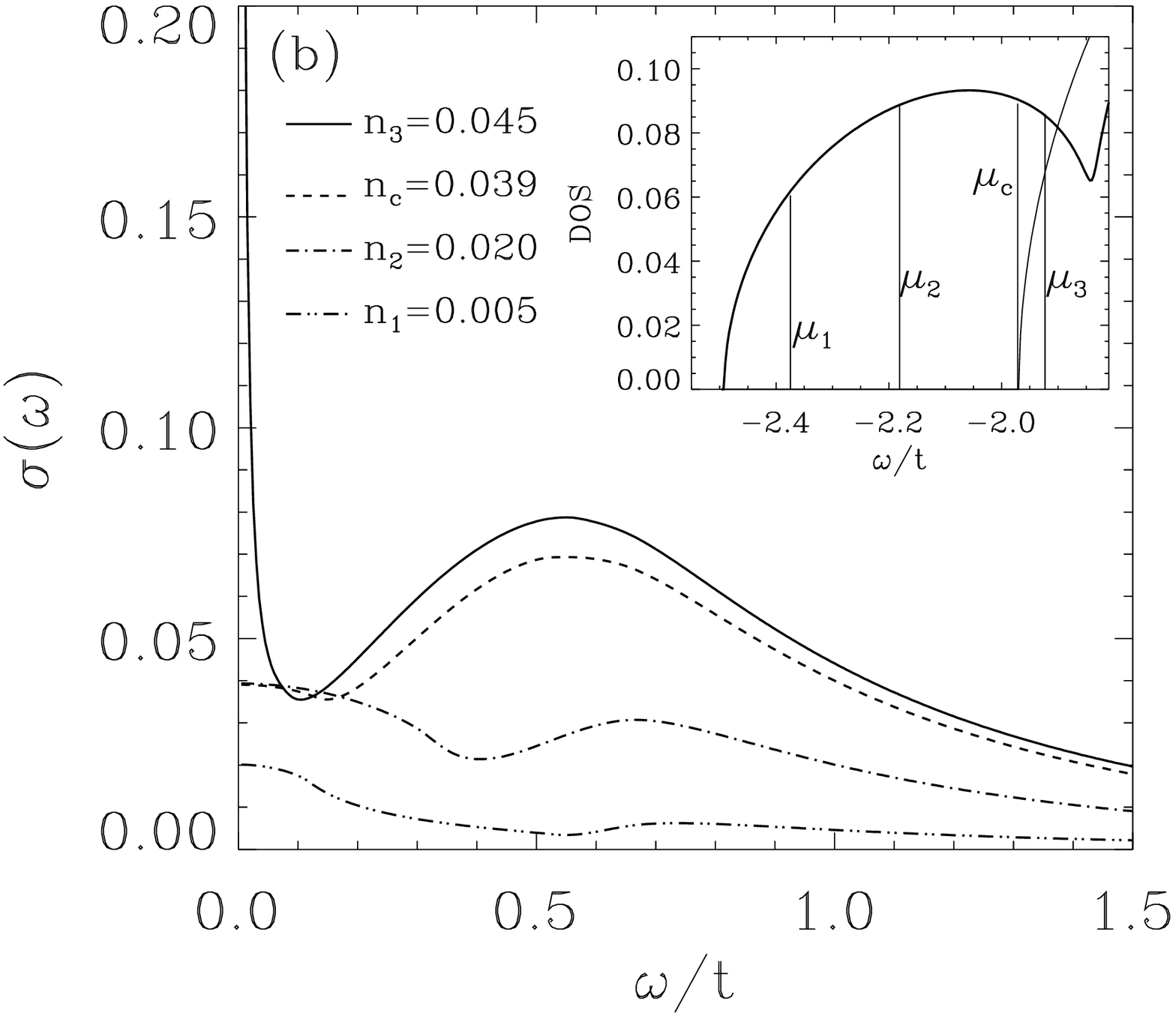}}
\caption{Density dependence of optical conductivity 
for (a) $J=1.0t$, $x=0.05$, and (b) $J=1.5t$, $x=0.05$ for $T=0K$. 
Inset shows the density of states
and the chemical potentials at zero temperature.
For $n>n_c$ the minority subband is occupied.}
\end{figure}

\noindent
additional sharp low frequency 
component is observed.

Fig. 4 shows the sensitivity of the predicted behavior to potential 
scattering. The main panel shows the evolution of the conductivity as 
the scalar part $W$ of the electron-impurity potential is varied. We 
study here a case in which at zero scalar potential the impurity band 
is well formed. As the potential is made more attractive, the impurity 
band features become more pronounced; as it is made more repulsive, 
the impurity band rapidly rejoins the main band and the extra feature 
in the 
conductivity is lost. The inset shows the dependence of the 
transition temperature on density at different potential strengths. 
We see that for attractive potential the dependence on potential 
strength is weak, but as the potential is made more repulsive the 
impurity band is destroyed and $T_c$ decreases.

It is instructive to compare the optical properties of DMS
to those of another system with strong carrier-spin 
couplings, namely CMR manganites such as
(La$_{1-x}$Ca$_x$)MnO$_3$ \cite{CMRReview}. In CMR,
instead of being dilute random impurities as in DMS,
the Mn ions form an ordered lattice. They 
possess large local moment, to which 

\begin{figure}
\epsfysize=2.3in
\centerline{\epsffile{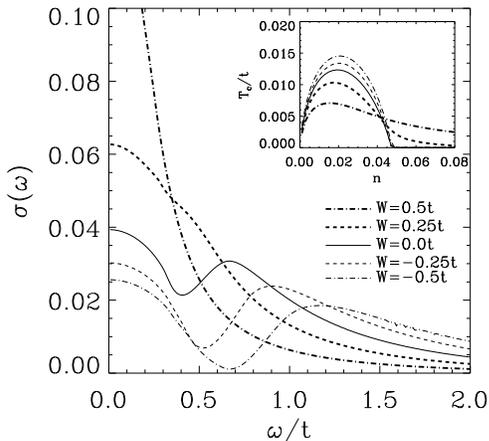}}
\caption{Optical conductivities for various of impurity potential scattering.
Inset shows the transition temperature as a function of density.}
\end{figure}

\noindent
mobile carriers are very 
strongly coupled. Thus instead of a spin-polarized impurity band, 
there is a spin-polarized conduction band, sufficiently 
well separated from other bands in the solid that the contribution to 
the optical conductivity arising from it may be clearly identified 
and analyzed \cite{Quijada98}. 
The periodic arrangement of the Mn sites means that (in the
absence of other physics) the scattering rate decreases as T is lowered
unlike in the DMS system. The conductivity turns out to be most usefully 
characterized by its integrated area (spectral weight), and the change 
of this quantity was shown to be a good predictor of the magnetic 
transition temperature \cite{Chattopadhyay00}.

These CMR ideas have limited applicability to the DMS systems.
For $J<1$ the physics is of extended states scattered by defects, and 
this is sufficiently different from the CMR situation that the spectral 
weight in DMS is not related in a useful manner to $T_c$
in this weak coupling situation. However, as $J$ is 
increased and the spin-polarized impurity band develops the physics 
becomes more analogous to that of CMR. For example, in Fig. 1(b) the 
two structures in the conductivity are relatively well separated, and 
the spectral weight in the lower feature increases as $T$ decreases 
below $T_c$. For $J>2$ (not shown) the low feature becomes
completely separated from the upper one. We therefore define an 
effective `impurity 
band spectral weight' for $J>1$ by integrating the conductivity from 
$\omega=0$ to the conductivity minimum.
We find again that the changes in 'impurity band spectral weight'
are very weak in DMS systems relative to those observed in CMR materials,
so that spectral weight ideas useful in CMR do not carry over to DMS.
This is an important qualitative distinction between DMS and CMR
materials.

To summarize, we have shown that the frequency, density, and temperature 
dependence of the conductivity contains important information about the 
physics of dilute magnetic semiconductors; in particular the formation 
of a spin-polarized impurity band leads to a peak centered at a 
non-zero frequency. We present several, at first sight, counterintuitive 
findings. We find an {\it increase} in scattering rate as $T$ is 
decreased, signalling enhanced carrier-spin coupling with increasing 
spin alignment. We also show that 
in certain doping and coupling regimes a very narrow conductivity 
peak could occur
corresponding to a slightly occupied minority spin band. It occurs 
because the repulsive interaction between local moments and 
``wrong-spin'' carriers suppresses the carrier amplitude at the 
impurity site, reducing the effective carrier-spin coupling 
and consequently narrowing the wrong-spin conductivity peak; 
in addition in three dimensions the vanishing of the density of states 
at the band edge further decreases the rate. This suggests that a small 
occupation of the minority band could be quite dangerous from the 
spintronic applications point of view, particularly in three
dimensional devices.

Finally, we note that the most interesting phenomena involve 
intermediate couplings, intermediate temperatures and non-zero 
frequency response. This regime is very difficult to treat by 
standard analytical or numerical methods; it is therefore fortunate 
that the dynamical mean field method allows access to this regime.

We thank H. D. Drew and A. Chattopadhyay for 
helpful conversations and 
DARPA(E.H.H. and S.D.S.), US-ONR (E.H.H and S.D.S.), and
the University of Maryland/Rutgers NSF-DMR-MRSEC (A.J.M.) for support.


\vspace{-0.5cm}

\begin{references}

\vspace{-1.5cm}

\bibitem{Ohno98}  H.\ Ohno, Science {\bf 281}, 951(1998); J.\ Magn.\ Magn.\
Mater. {\bf 200}, 110(1999).

\bibitem{Ohno96}  H.\ Ohno {\it et al.,}
Appl.\ Phys.\ Lett {\bf 69}, 363 (1996); A.\ Van Esch et al., \prb
{\bf 56}, 13103(1997).

\bibitem{Matsukura98}  F. Matsukura {\it et al.}, 
\prb {\bf 57}, R2037(1998).

\bibitem{Reed01} M. Reed {\it et al.}, Appl. Phys. Lett. {\bf 79}, 3473 
(2001); N. Theodoropoulou {\it et al.}, cond-mat/0201492 (2002).

\bibitem{Akai98}  H.\ Akai, Phys.\ Rev.\ Lett {\bf 81}, 3002(1998).

\bibitem{Dietl00}  T. Dietl {\it et al.}, Science {\bf 287}, 1019(2000).

\bibitem{Litvinov01}  V.\ I.\ Litvinov and V.\ K.\ Dugaev, Phys.\ Rev.\
Lett. {\bf 86}, 5593(2001).

\bibitem{Dietl02} T. Dietl, cond-mat/0201282;
J.\ K\"{o}nig {\it et al.}, cond-mat/0111314. 

\bibitem{Bhatt00} M.\ Berciu and R. N. Bhatt, \prl {\bf 87},
7203 (2000).

\bibitem{Nagaev96} A. Kaminski and S. Das Sarma, cond-mat/0201229 (2002);
C. Timm {\it et al}., cond-mat/0201411 (2002).

\bibitem{Chattopadhyay01a}  A. Chattopadhyay {\it et al.}, 
\prl {\bf 87}, 227202-1 (2001).

\bibitem{dassarma01} S. Das Sarma, Am. Scientist {\bf 89}, 516 (2001).

\bibitem{Quijada98}  M. Quijada {\it et al.}, 
\prb {\bf 58}, 16093 (1998).

\bibitem{CMRReview} See e.g.  \textit{Colossal
Magnetoresistive Oxides}, Y. Tokura, ed. (Gordon and Breach, 1999).


\bibitem{Kotliar}A. Georges {\it et al.}, Rev. Mod. Phys {\bf 68}, 13 (1996).

\bibitem{Sanvito01}  S.\ Sanvito, {\it et al.}, \prb {\bf 63}, 165206(2001).

\bibitem{Chattopadhyay00}  A. Chattopadhyay {\it et al.}, 
\prb {\bf 61}, 10738 (2000).
\end{references}
\end{document}